\documentclass{article}

\usepackage{authblk}
\usepackage[utf8]{inputenc}
\usepackage[T1]{fontenc}
\usepackage{lmodern}
\usepackage{amsmath,amssymb,amsfonts}
\usepackage{bm}
\usepackage{geometry}
\usepackage{graphicx}
\usepackage{microtype}
\usepackage{subcaption}
\usepackage{hyperref}
\usepackage[
  backend=biber,
  style=numeric,        
  sorting=none          
]{biblatex}
\begin{document}

\title{Can Stochastic Clocks in FLRW Minisuperspace Prevent Dynamical Singularities?}

\author[1]{Pradosh Keshav MV\thanks{Email: pradosh.keshav@res.christuniversity.in}}
\affil[1]{Department of Physics and Electronics, Christ University, Bangalore, India-560029}

\date{}

\maketitle

\begin{abstract}
We develop a stochastic extension of the Wheeler–DeWitt equation in FLRW minisuperspace and show that quantum backreaction can dynamically regulate the big bang singularity without imposing external boundary conditions. Using Laplace–Beltrami quantisation and an open-system treatment of coarse-grained graviton modes, we obtain a stochastic Hamiltonian evolution equation in which the diffusion coefficient takes the form $\sigma(a)\propto a^2$. This multiplicative noise vanishes at the origin and renders $a=0$ an entrance boundary in Feller’s classification, leading to super-exponential suppression of the Laplace–Beltrami weighted stationary density and zero probability flux into the singular point. At large scale factor, the global behaviour depends on the cosmological sector: de Sitter and positive potential-dominated regimes exhibit power-law stationary tails, whereas confining potentials or negative effective cosmological constant lead to an entrance boundary at infinity and a globally normalizable steady state. Taken together, these results indicate that stochastic backreaction arising from semiclassical coarse-graining provides a consistent and dynamical mechanism for singularity avoidance in minisuperspace quantum cosmology.
\end{abstract}
\textbf{Keywords:} Stochastic clocks, Wheeler--DeWitt equation, Quantum cosmology

\section{Introduction}

A central question in quantum cosmology is whether the big-bang singularity can be resolved dynamically, without imposing external boundary conditions \cite{ashtekar2006quantum, hartle1983wave}. In canonical Wheeler–DeWitt (WD) quantisation, the vanishing of the wavefunction at $a=0$ is usually enforced via the DeWitt condition \cite{dewitt1967quantum, dewitt1967quantum2}, while loop quantum cosmology and related polymer quantisations achieve resolution through modified dynamics \cite{isham1993canonical,kuchavr2011time,date2004effective}. More recently, relational-time models \cite{nambu2022qubit,rotondo2019clock} and stochastic formulations of gravity \cite{erlich2018stochastic} have suggested that quantum backreaction may itself drive singularity avoidance; however, in most treatments, the stochastic structure is postulated rather than derived. What remains open is whether a consistent minisuperspace derivation can generate stochastic dynamics that regulate the singular boundary.

In this Letter, we provide such a derivation, extending our earlier work on stochastic clocks \cite{mv2025relational}. For definiteness, we consider a homogeneous and isotropic universe with lapse $N(t)$, scale factor $a(t)$, cosmological constant $\Lambda$, and homogeneous scalar clock $\phi(t)$. In Arnowitt–Deser–Misner (ADM) variables, the Einstein–Klein–Gordon action reduces to the standard minisuperspace form \cite{misner1969quantum, kiefer1988wave}:
\begin{equation}
S=2\pi^2\!\int\!dt\,N\Big[-\frac{3a\dot a^2}{N^2}+3ka-\Lambda a^3+\frac{a^3\dot\phi^2}{2N^2}-a^3 V(\phi)\Big],
\label{eq:action}
\end{equation}
with canonical momenta
\begin{equation}
p_a=-12\pi^2\frac{a\dot a}{N},\qquad p_\phi=2\pi^2\frac{a^3\dot\phi}{N},
\label{eq:momenta}
\end{equation}
and Hamiltonian constraint
\begin{equation}
H_0=-\frac{p_a^2}{24\pi^2 a}-2\pi^2(3ka-\Lambda a^3)+\frac{p_\phi^2}{4\pi^2 a^3}+2\pi^2 a^3 V(\phi)=0.
\label{eq:constraint}
\end{equation}
Fixing the relational gauge by identifying the homogeneous scalar field with a relational time parameter $\phi\equiv \varphi$, and choosing \(\varphi =t\) implies $\dot\phi=1$ and hence $N=2\pi^2 a^3/p_\phi$. For completeness, a brief derivation of the minisuperspace action from four-dimensional ADM Einstein–Hilbert action is provided in Appendix \ref{app:a1}. 

Solving the constraint \eqref{eq:constraint} for $p_\phi$ then yields the effective Hamiltonian generating evolution in the scalar clock,
\begin{equation}
p_\phi \equiv H_{\rm eff}(a,p_a;\phi)
= \sqrt{\frac{a^2 p_a^2}{6}
+8\pi^4 a^4\Big[3k -\big(\Lambda+V(\phi)\big)a^2\Big]}\,.
\label{eq:Heff}
\end{equation}
We explicitly take the positive branch $H_{\rm eff}>0$ in order to define a forward evolution in the clock variable. The square-root form is only valid in the classically allowed region of the $(a,p_a)$ phase space, where the argument of the square root is non-negative,
\begin{equation}
\frac{a^2 p_a^2}{6}
+8\pi^4 a^4\Big[3k -\big(\Lambda+V(\phi)\big)a^2\Big]\ge 0,
\end{equation}
while outside this region $p_\phi$ becomes imaginary and the dynamics correspond to a Euclidean (tunnelling) sector that we do not consider further in this Letter.

From this reduced starting point, we implement Laplace–Beltrami (LB) quantisation, which replaces the naive substitution $p_a \mapsto -i\hbar\,\partial_a$ with the covariant operator 
\[
\hat{T} = -\frac{\hbar^2}{2}\nabla_{\mathrm{LB}} = -\frac{\hbar^2}{2\sqrt{G}}\partial_a\!\left(\sqrt{G}G^{aa}\partial_a\right),
\]
ensuring correct operator ordering and self-adjointness with respect to the minisuperspace DeWitt metric.\footnote{The Laplace–Beltrami operator $\nabla_{\mathrm{LB}}$ is the natural generalisation of the Laplacian to curved configuration spaces and guarantees hermiticity under the WD inner product.} This yields a modified WD equation whose imaginary part, following the Feynman–Vernon influence functional, leads to a stochastic Wheeler–DeWitt (sWD) equation with multiplicative stochastic backreaction \cite{hu1994quantum, hu1995back, hu2008stochastic}.

In the semiclassical regime, the sWD equation reduces to an effective stochastic differential equation (SDE) for the scale factor $a(\phi)$ of the form $da = f(a)\,d\phi + \sigma(a)\circ dW_\phi$, where the circle denotes the Stratonovich interpretation.\footnote{The Stratonovich calculus preserves ordinary chain rules and geometric covariance, while the Itô formulation modifies the drift by an additional noise-induced term $f(a)\mapsto f(a)+\tfrac{1}{2}\sigma(a)\sigma'(a)$. We employ Stratonovich throughout for compatibility with the canonical and geometric structure of minisuperspace.}  This allows us to characterise the small- and large-$a$ behaviour using Feller’s boundary classification,\footnote{Feller’s classification determines whether a boundary (here $a=0$ or $a\to\infty$) is accessible, reflecting, absorbing, natural, or entrance, based solely on the asymptotics of the drift $f(a)$ and diffusion $\sigma(a)$.} together with the stationary and transient behaviour of the associated Fokker–Planck (FP) equation governing the probability density of the process \cite{feller1954diffusion, risken1989fokker}. Within this framework, the question of whether stochastic backreaction dynamically regulates the classical singular boundary at $a=0$ becomes precise: one asks whether $a=0$ becomes an entrance (non-accessible) boundary or remains reachable with non-zero probability.

The remainder of the Letter develops this construction, combining analytic asymptotic criteria with numerical evolution of the stochastic dynamics, to assess whether multiplicative stochasticity can generically suppress the classical singular behavior and determine the global fate of minisuperspace trajectories at large $a$.

\section{Quantization and (semi-classical) stochastic reduction}

The first step is to fix the factor ordering and inner product so that the minisuperspace dynamics are well-defined. Since the reduced kinetic term in the Hamiltonian takes the form $p_a^2/(24\pi^2 a)$, the corresponding one-dimensional DeWitt metric on configuration space is \(G_{aa}=24\pi^2 a \equiv c\,a\), with constant $c=24\pi^2$. The LB prescription yields the kinetic operator
\begin{equation}
\widehat T_a=-\frac{\hbar^2}{2}\,\Delta_{\rm LB}
=-\frac{\hbar^2}{2}\frac{1}{\sqrt{G_{aa}}}\,\partial_a\!\Big(\sqrt{G_{aa}}\,G^{aa}\,\partial_a\Big)
=-\frac{\hbar^2}{2c}\,a^{-1/2}\partial_a\!\big(a^{-1/2}\partial_a\big).
\label{eq:LB}
\end{equation}
In what follows, we absorb the overall constant $c$ into a rescaling of the canonical variables; this affects only the normalization of the kinetic term, not the relative scaling with $a$ nor the resulting stochastic structure. With this convention, the kinetic operator takes the compact form
\[
\widehat T_a=-\frac{\hbar^2}{2}\,a^{-1/2}\partial_a\!\big(a^{-1/2}\partial_a\big).
\]

The resulting Wheeler--DeWitt inner product is uniquely determined by the metric measure
\begin{equation}
\langle\Phi|\Psi\rangle=\int_0^\infty da\,\mu(a)\,\Phi^\ast\Psi,\qquad 
\mu(a)=\sqrt{G_{aa}}\propto a^{1/2}.
\label{eq:measure}
\end{equation}
Self-adjointness of $\widehat T_a$ requires that the boundary form vanish at $a=0$, which is enforced by the standard LB reflecting condition \cite{kiefer2010avoidance}
\begin{equation}
\partial_a\!\big(\sqrt{\mu(a)}\,\Psi(a,\phi)\big)\Big|_{a=0}=0,
\label{eq:LBreflect}
\end{equation}
such that there is zero probability flux into the classical singularity. Promoting $p_a\to -i\hbar\,\partial_a$ and $p_\phi\to -i\hbar\,\partial_\phi$ in the corrected effective Hamiltonian \eqref{eq:Heff} finally yields a Schrödinger-type evolution equation in the relational time $\phi$:
\begin{equation}
i\hbar\,\partial_\phi\Psi(a,\phi)
=H_{\rm eff}\big(a,\widehat p_a;\phi\big)\Psi(a,\phi).
\label{eq:SchrClock}
\end{equation}

Having established the above operator structure, we now introduce quantum backreaction in a controlled way. Fluctuations around the minisuperspace trajectory generate an influence functional of the form \cite{calzetta1994noise, calzetta1999stochastic}
\begin{equation}
\mathcal{F}[a]=\exp\!\Big\{\frac{i}{2}\!\int\!da\,da'\,O(a)\,G(a,a')\,O(a')\Big\},
\label{eq:IF}
\end{equation}
where $G$ is the retarded kernel of the quadratic fluctuation operator and $O(a)$ is the coupling of the homogeneous scalar to geometry. A natural geometric choice is
\begin{equation}
O(a)=\alpha\,a^2 V(\phi),
\label{eq:Oa}
\end{equation}
which vanishes as $a\to 0$. This behavior effectively suppresses stochastic corrections at what would otherwise be a singularity, since the scalar couples through \(a^3 V(\phi)\) in the action, leading the quadratic correction to scale as \(a^2 V(\phi)\). By performing a Hubbard–Stratonovich transformation, we linearize equation \eqref{eq:IF} and introduce a Gaussian noise term \(\xi(a)\) with the correlator $\langle\xi(a)\xi(a')\rangle=\mathcal N\,G(a,a')$. The result is the stochastic Wheeler--DeWitt equation
\begin{equation}
i\hbar\,\partial_\phi\Psi=\Big(\widehat H_{\rm eff}+\xi(a)\,O(a)\Big)\Psi,
\label{eq:sWD}
\end{equation}
which we interpret in the Stratonovich sense (see \cite{yhu2009stochastic, yhu2019some}), consistent with its path-integral origin and reparametrisation covariance. In the narrow-tube (short-memory) approximation $G(a,a')\approx\delta(a-a')$, the noise strength becomes local, $\zeta\propto\mathcal N\alpha^2 V(\phi)^2$ \cite{calzetta2001coarse, arnold2000symmetric}.

The semiclassical reduction proceeds by inserting a WKB ansatz $\Psi=A\,e^{i\mathcal S/\hbar}$ with slowly varying amplitude. At leading order, \eqref{eq:sWD} reduces to a noisy Hamilton--Jacobi equation that is equivalent to the Stratonovich stochastic differential equation
\begin{equation}
\frac{da}{d\phi}=v(a)+\sigma(a)\circ dW_\phi,
\qquad \sigma(a)=\sqrt{\zeta}\,a^2,
\label{eq:SDE}
\end{equation}
where $W_\phi$ is a Wiener process and the deterministic drift $v(a)$ is fixed by the classical constraint, behaving as $v(a)=c\,a+\mathcal O(a^3)$ for regular $V(\phi)$ and $\Lambda$ as $a\to 0$. Here, transforming \eqref{eq:SDE} to Itô form adds the spurious drift $\tfrac12\sigma\sigma'$, leading to the FP equation
\begin{equation}
\partial_\phi P=-\partial_a\!\big(f(a)P\big)+\tfrac12\,\partial_a^2\!\big(\sigma^2(a)P\big),\qquad f(a)=v(a)+\tfrac12\sigma\sigma',
\label{eq:FP}
\end{equation}
with small-$a$ asymptotics
\begin{equation}
f(a)=c a+\zeta a^3+\mathcal O(a^5),\qquad 
\sigma^2(a)=\zeta a^4.
\label{eq:FPsmall}
\end{equation}

This construction achieves two goals. First, it provides a unique quantisation scheme free of factor-ordering ambiguities and equipped with a well-defined Hilbert space measure. Second, it derives a stochastic extension of the Wheeler--DeWitt equation directly from the influence functional, showing that the resulting multiplicative noise naturally vanishes at the singularity. The system–environment decomposition, the construction of the influence functional, and the semiclassical reduction leading to the effective stochastic dynamics are detailed in Appendices~\ref{app:a2} and \ref{app:a3}.

\section{Stochastic boundary analysis in minisuperspace}

We now present the central result of this work: in the sWD framework, the classical big-bang singularity at $a=0$ is dynamically avoided, without imposing external boundary conditions of the DeWitt type. The avoidance arises from the precise interplay between two features of the stochastic minisuperspace dynamics: the multiplicative diffusion $\sigma(a)\propto a^2$ which degenerates at $a=0$, and the regular small-$a$ relational drift $v(a)=c\,a+\mathcal O(a^3)$ (see Appendix~\ref{app:a4} for a classification of more general diffusion scalings). Together, these define the Stratonovich
process (converted below to Itô form)
\begin{equation}
da=f(a)\,d\phi+\sigma(a)\,dW,\qquad
f(a)=v(a)+\tfrac12\sigma(a)\sigma'(a),\qquad
\sigma(a)=\sqrt{\zeta}\,a^2,
\label{eq:Ito-corrected}
\end{equation}
and the associated FP equation for the flat-measure density \(P(a,\phi)\) (with
\(D(a)=\tfrac12\sigma^2(a)\)). In the small-\(a\) regime
\begin{equation*}
f(a)=c a+\zeta a^3+\mathcal O(a^5),\qquad D(a)=\tfrac{\zeta}{2}a^4.
\end{equation*}

At this point, it is convenient to determine the reachability of \(a=0\) using Feller's boundary classification. We define the scale function \(s(a)\) and the speed density \(m(a)\) (following the conventions of Gardiner \cite{gardiner2004quantum}):
\begin{equation}
s(a)=\exp\!\Big[-\int^a \frac{2f(x)}{\sigma^2(x)}\,dx\Big],\qquad
m(a)=\frac{2}{\sigma^2(a)\,s(a)}.
\end{equation}
With \(\sigma^2(x)=\zeta x^4\) and \(f(x)=c x+\zeta x^3+\cdots\) the integrand evaluates to
\begin{equation*}
\frac{2f(x)}{\sigma^2(x)}=\frac{2}{x}+\frac{2c}{\zeta x^3} +\mathcal O(1),
\end{equation*}
so that (integrating explicitly)
\begin{equation*}
\int^a \frac{2f(x)}{\sigma^2(x)}\,dx = 2\ln a -\frac{c}{\zeta a^2} + C,
\end{equation*}
and therefore the scale and speed functions take the form
\begin{equation}
s(a)=C_1\,a^{-2}\,e^{\,c/(\zeta a^2)},\qquad
m(a)=\frac{2}{\zeta C_1}\,a^{-2}\,e^{-\,c/(\zeta a^2)},
\label{eq:s-m-correct}
\end{equation}
with \(C,C_1\) integration constants (absorbed in normalization below). The leading small-\(a\) behaviour
is determined by the exponential factors: as \(a\to 0\),
\begin{equation}
s(a)\sim a^{-2} e^{\,c/(\zeta a^2)}\to +\infty,\qquad
m(a)\sim a^{-2} e^{-\,c/(\zeta a^2)}\to 0.
\end{equation}
Hence, the integrals at the origin satisfy
\(\displaystyle\int_0 s(a)\,da = +\infty\) while \(\displaystyle\int_0 m(a)\,da <\infty\). According to Feller's criterion, this combination corresponds to an \emph{entrance} boundary at \(a=0\), i.e., trajectories started at any \(a>0\) almost surely do not reach \(a=0\) in finite relational time \(\phi\) (see e.g. \cite{hu2021weyl,karlin2014first}). (Physically, however, an entrance boundary is inaccessible from the interior, although it may serve as an initial boundary, but is not attained dynamically.) 

A straightforward corollary is that the probability current
\(
J=fP-\partial_a(DP)
\)
vanishes at \(a=0\) \cite{risken1996fokker}. Including the LB weight \(\mu(a)\propto a^{1/2}\) (used to define the physical density \(\mathcal P=\mu P\)), one finds that the LB-weighted probability density vanishes at the origin:
\begin{equation*}
\lim_{a\to 0}\mathcal P(a,\phi)=\lim_{a\to 0}\mu(a)P(a,\phi)=0,
\end{equation*}
where the probability flux into the singularity is dynamically forbidden.

\begin{figure*}
  \centering
  \begin{subfigure}[t]{0.48\textwidth}
    \centering
    \includegraphics[width=\linewidth]{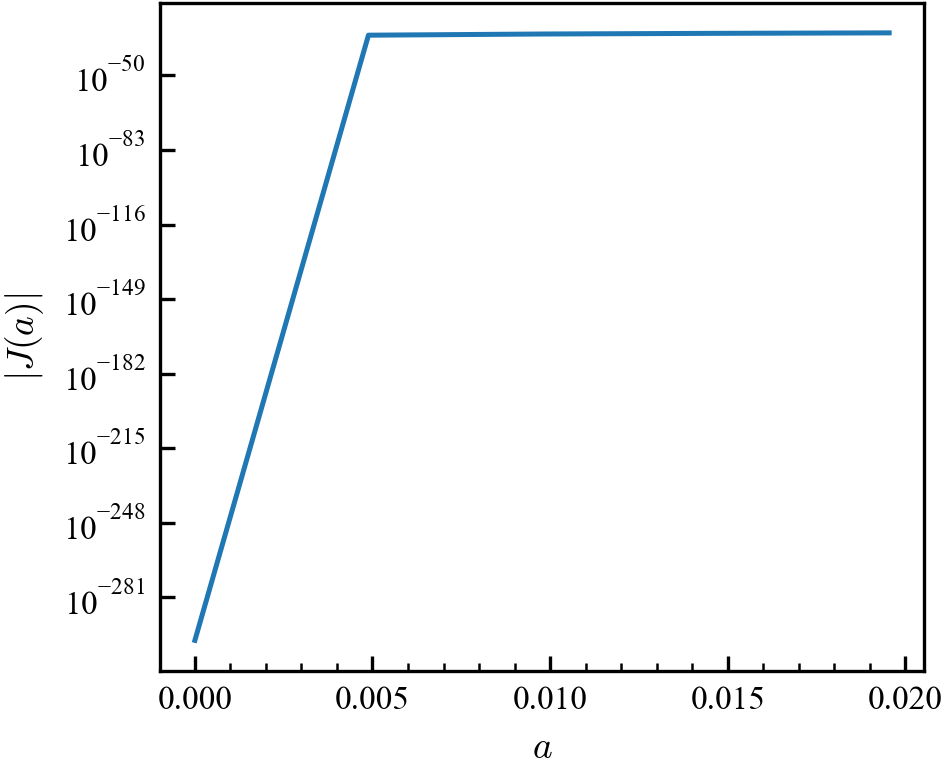} 
    \caption{Probability current near $a=0$} 
  \end{subfigure}
  \hfill
  \begin{subfigure}[t]{0.48\textwidth}
    \centering
    \includegraphics[width=\linewidth]{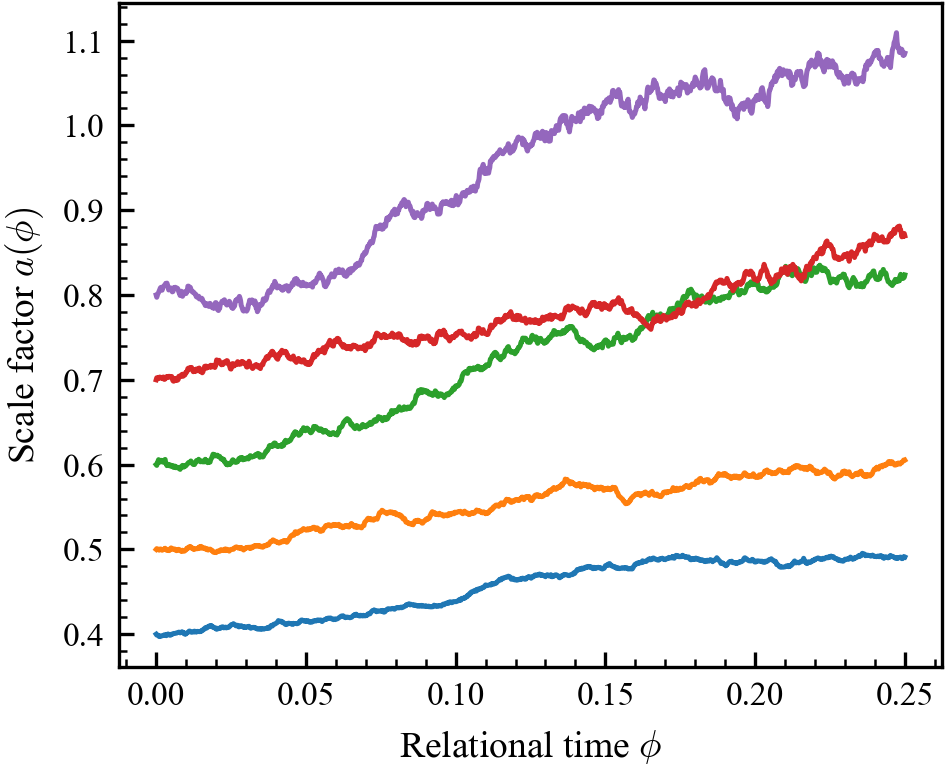} 
    \caption{Stochastic trajectories} 
  \end{subfigure}

  \caption{\small Boundary dynamics at the singularity in the sWD framework. (a) Probability current $J(a)=fP-\partial_a(DP)$ computed from the stationary solution with zero-flux boundary conditions tends to zero as $a\to0$,  verifying the entrance classification of the boundary. (b) Representative Stratonovich-consistent stochastic trajectories $a(\phi)$ (Heun scheme) started at $a>0$ and never reach the singularity, making \(a=0\) dynamically inaccessible.}
  \label{fig:boundary_dynamics}
\end{figure*}

Since the stochastic dynamics is formulated in the Stratonovich sense, converting to the It\^o form introduces the standard noise-induced drift term $\tfrac12\sigma\sigma'=\zeta a^3$. The effective drift entering the FP equation is therefore
\[
f(a)=v(a)+\tfrac12\sigma\sigma'=c a+\zeta a^3+\mathcal{O}(a^5).
\]This argument can be reinforced by the stationary (zero-current) solution of the FP equation. Using the integral above, the zero-current condition \(fP_*-\partial_a(DP_*)=0\) yields
\begin{equation}
P_*(a)\propto \frac{1}{D(a)}\exp\!\Big(\int^a \frac{2f(x)}{\sigma^2(x)}\,dx\Big)
=\frac{2}{\zeta}\,a^{-2}\,e^{-\,c/(\zeta a^2)}.
\label{eq:stationary-correct}
\end{equation}After multiplication by the LB weight \(\mu(a)\propto a^{1/2}\) the physical density behaves as
\begin{equation}
\mathcal P_*(a)\propto a^{-3/2}\,e^{-\,c/(\zeta a^2)}\xrightarrow{a\to 0}0,
\end{equation} i.e., the origin is suppressed super-exponentially, and the stationary distribution is integrable at the origin. This confirms that probability flux into \(a=0\) is zero and that singularity avoidance is realised dynamically by the multiplicative noise structure (as illustrated in Figure \ref{fig:boundary_dynamics} ).

\paragraph{The white-noise (Markov) approximation:} 
In the full theory, the stochastic deformation arises from integrating out coarse-grained transverse–traceless (TT) graviton perturbations, which act as an environment for the minisuperspace degrees of freedom. In general, this procedure generates a nonlocal (coloured) noise kernel $G(\phi,\phi')$, reflecting finite correlation time and memory effects in the gravitational sector \cite{hu1990quantum}. The Markov, or white-noise, approximation becomes valid when the characteristic graviton correlation time $\tau_{\rm grav}$ is much shorter than the timescale $\tau_{\rm clock}$ over which the relational variables evolve,
\begin{equation*}
\tau_{\rm grav} \ll \tau_{\rm clock}\quad\Rightarrow\quad
G(\phi,\phi')\simeq 2\zeta\,\delta(\phi-\phi'),
\end{equation*}
where the Dirac delta shows the instantaneous decorrelation of environmental modes. Physically, this limit corresponds to the standard Markovian in open-system theory in which unresolved UV graviton modes have support at frequencies much larger than the characteristic inverse evolution time of the clock \cite{accardi2000white}. An explicit model calculation demonstrating how the coarse-grained TT graviton sector produces a local noise kernel in this regime is given in Appendix~\ref{app:B_markov}.

If the gravitational noise were weakly coloured rather than strictly white, its primary effect would be to renormalize $\zeta$ and introduce small memory corrections. Importantly, the super-exponential suppression of diffusion near $a\to0$ remains robust under such perturbations, since the leading behaviour is controlled by the multiplicative factor $\sigma(a)\propto a^2$. Dissipative contributions are likewise subleading in the slow-clock regime considered here and do not modify the entrance character of the boundary at $a=0$. Only for very long-range temporal correlations—where the memory kernel decays more slowly than any power—could the boundary classification or infrared stationary behaviour change; such genuinely non-Markovian extensions are beyond the scope of the present Letter.

\begin{figure*}[t]
  \centering
  \begin{subfigure}[t]{0.48\textwidth}
    \centering
    \includegraphics[width=\linewidth]{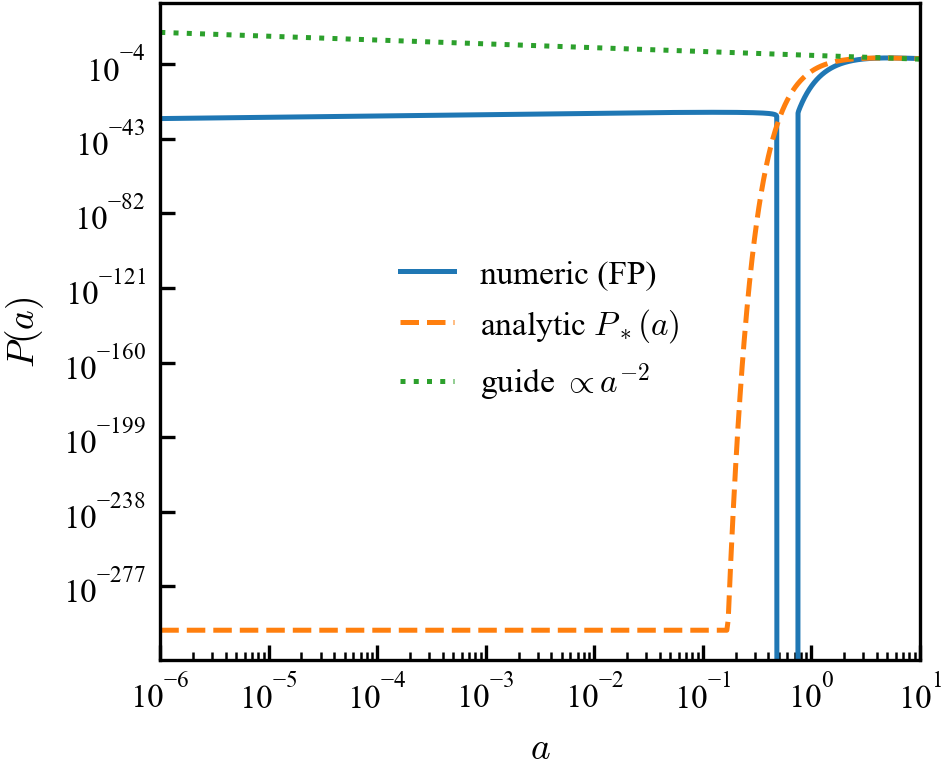} 
    \caption{Stationary density (flat measure)}  
  \end{subfigure}
  \hfill
  \begin{subfigure}[t]{0.48\textwidth}
    \centering
    \includegraphics[width=\linewidth]{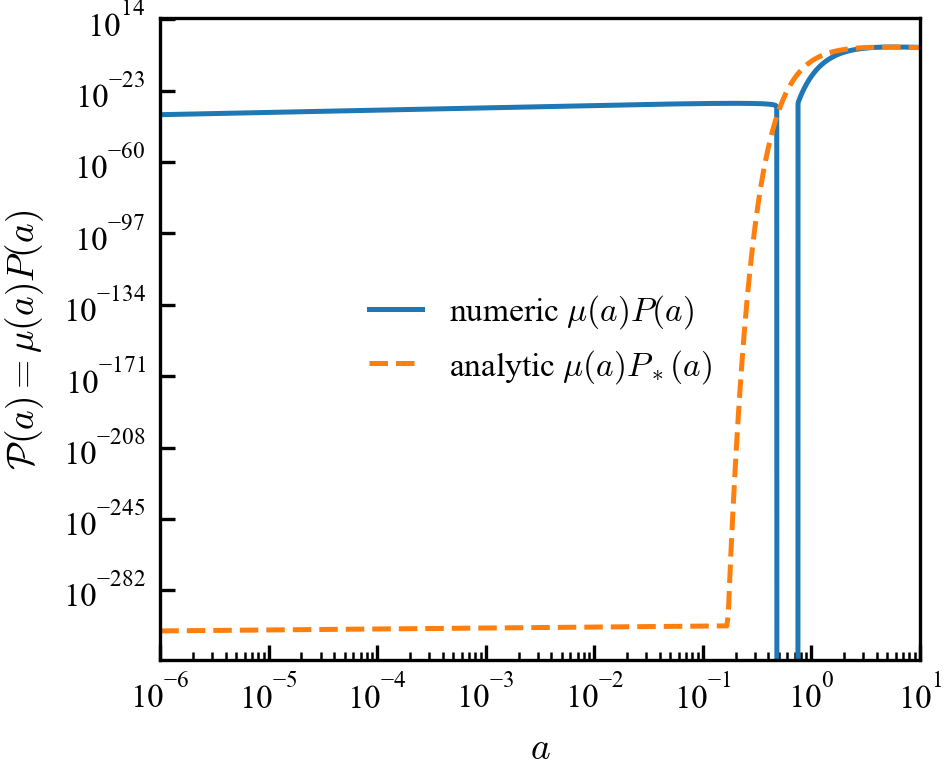} 
    \caption{LB-weighted stationary density}  
  \end{subfigure}

  \caption{\small Stationary probability density in the sWD minisuperspace with zero-flux boundary conditions at both ends. (a) Flat-measure stationary density $P(a)$ from the finite-volume  FP solver (solid) agrees with the analytic zero-current solution (dashed), and follows an $a^{-2}$ tail at large $a$ (dotted guide). (b) Laplace–Beltrami weighted density $\mathcal{P}(a)=\mu(a)P(a)$ exhibits super-exponential suppression as $a\to 0$, confirming that the singularity is dynamically inaccessible, while decaying at large $a$ according to the analytic profile.}
  \label{fig:stationary_density}
\end{figure*}

\paragraph{Numerical reproducibility:} The numerical FP and stochastic trajectory results shown in Figures \ref{fig:boundary_dynamics} and \ref{fig:stationary_density} were obtained with a conservative finite-volume solver on $(a\in[a_{\min},a_{\max}])$ with $(a_{\min}=10^{-6})$, $(a_{\max}=10)$, using a uniform grid of $(N=2048)$ points. Implicit backward-Euler time stepping was employed for the FP equation with step size $(\Delta\phi=10^{-3})$, iterated until convergence at tolerance $(10^{-10})$. Zero-flux boundary conditions were imposed at both $(a_{\min})$ and $(a_{\max})$ to match the analytic zero-current stationary solution. The LB weighting was included in post-processing. Stochastic trajectories were generated with a Stratonovich-consistent Heun scheme (predictor–corrector with Wiener increments), using step size $(\Delta\phi=2\times 10^{-4})$ over intervals up to $(\phi=0.25)$. Parameter values used in the figures are $(c=1.0)$ and $(\zeta=0.05)$; other choices give qualitatively identical suppression and boundary behaviour.

\section{Global asymptotics and normalisability}

The near-singularity analysis showed that $a=0$ is universally an entrance boundary with super-exponential suppression of the LB-weighted density, independent of the details of $(V(\phi),\Lambda,k)$. To establish global consistency, we must also examine the opposite limit $a\to\infty$, where the classification depends sensitively on the cosmological sector.

Using the transport coefficients of Eq.~\eqref{eq:Ito-corrected} with $f(a)=v(a)+\zeta a^3$ and $D(a)=\tfrac{\zeta}{2}a^4$, together with the corrected Hamiltonian constraint
\begin{equation}
\label{eq:constraint_letter}
p_\phi^2=\frac{a^2 p_a^2}{6}
+24\pi^4 k a^4
-8\pi^4 a^6\big[\Lambda+V(\phi)\big],
\end{equation}
we analyze the large-$a$ drift $v(a)=\partial H_{\rm eff}/\partial p_a = a^2 p_a/(6p_\phi)$ in different cosmological regimes. The square-root form of $H_{\rm eff}$ implies that the radicand in \eqref{eq:constraint_letter} must remain non-negative; this constraint determines the asymptotic scaling of $p_a$ and $p_\phi$ and hence of $v(a)$.

\paragraph{de Sitter / positive vacuum-energy dominated regime ($\Lambda_{\rm tot}:=\Lambda+V(\phi)>0$).}

When the total vacuum energy $\Lambda_{\rm tot}$ is positive, the $a^6$ term in \eqref{eq:constraint_letter} is negative and must be balanced by the kinetic contribution in order for $p_\phi^2$ to remain non-negative at large $a$. In the asymptotic regime, this requires
\begin{equation}
\frac{a^2 p_a^2}{6} \simeq 8\pi^4 \Lambda_{\rm tot}\,a^6 \quad\Rightarrow\quad
p_a^2 \simeq 48\pi^4 \Lambda_{\rm tot}\,a^4,
\end{equation}
so that $p_a \propto a^2$ while the curvature term $24\pi^4 k a^4$ contributes only at subleading order. Consequently, $p_\phi^2$ grows at most like $a^4$ and the ratio $p_a/p_\phi$ approaches a constant. It follows that the deterministic drift behaves as
\begin{equation}
v(a)=\frac{a^2 p_a}{6p_\phi}=\mathcal{O}(1)
\qquad (a\to\infty),
\end{equation}
while the multiplicative noise term in $f(a)=v(a)+\zeta a^3$ grows as $a^3$. Thus, for sufficiently large $a$, the $\zeta a^3$ term dominates,
\[
f(a)\simeq \zeta a^3,
\]
and the ratio entering the Feller analysis becomes
\[
\frac{2f}{\sigma^2}=\frac{2}{a}+\mathcal{O}(a^{-4}),\qquad 
s(a)\sim a^{-2},\qquad m(a)\sim \frac{2}{\zeta a^2}.
\]
Therefore $a=\infty$ is a \emph{regular} boundary, and the stationary solution (cf.\ Eq.~\eqref{eq:stationary-correct}) asymptotes to
\[
P_*(a)\sim \frac{2}{\zeta a^2},\qquad 
\mathcal P_*(a)\sim \frac{2}{\zeta}a^{-3/2},
\]
which is integrable at infinity since $\int^\infty a^{-3/2}da$ converges.

\paragraph{Negative cosmological sector / confining large-$a$ drift.}

If the combined cosmological term is effectively negative ($\Lambda_{\rm eff}:=\Lambda+V(\phi)<0$) at large $a$ (for instance, due to a negative bare cosmological constant or a sufficiently steep negative potential), then the $a^6$ contribution in \eqref{eq:constraint_letter} is positive and dominates. In such cases $p_\phi^2 \sim 8\pi^4|\Lambda_{\rm tot}|\,a^6$ while the kinetic and curvature contributions grow more slowly (as $a^4$ or less). This allows for a strongly restoring deterministic drift. More generally, one may parametrise the asymptotic behaviour of the drift as
\begin{equation}
f(a)\sim -\gamma a^p,\qquad \gamma>0,
\end{equation}
with $p>3$ corresponding to a genuinely confining large-$a$ sector. In this regime the competition between drift and diffusion gives
\[
\frac{2f}{\sigma^2}\sim -\frac{2\gamma}{\zeta}a^{\,p-4}\qquad (p>4),
\]
so that
\[
\int^a \frac{2f}{\sigma^2}dx \to -\infty\quad\text{as }a\to\infty,
\]
and $a=\infty$ is an \emph{entrance} boundary. The stationary density then decays faster than any power, resulting in a globally normalisable distribution. For the marginal case $p=4$ one finds an exponential tail, which is still normalisable, while for $3<p<4$ the tail remains power-law but with a stronger suppression than in the de Sitter-like case.

\medskip

Thus, while dynamical singularity avoidance at $a=0$ is independent of model details, the global behaviour of the stationary distribution depends on the large-scale cosmological sector \cite{starobinsky1994equilibrium}. Positive total vacuum-energy (de Sitter–like) regimes lead to a regular boundary at infinity with a power-law stationary tail, whereas sufficiently confining or negative-cosmological sectors render $a=\infty$ an entrance boundary and yield global normalisability. This dichotomy emphasises that stochastic backreaction removes the singularity universally, but the infrared completion of the probability measure is controlled by cosmological input encoded in the asymptotic form of $f(a)$.

\section{Outlook}

The central result of this Letter is that, within the sWD framework and under the assumptions of minimal coupling, LB operator ordering, and the Markov limit for coarse-grained graviton modes, the classical singularity at $a=0$ becomes dynamically inaccessible. In this setting, the multiplicative diffusion $\sigma(a)=\sqrt{\zeta}\,a^2$, obtained from the minisuperspace influence functional, combines with the regular drift $v(a)=c\,a$ to place $a=0$ in Feller’s entrance class. Analytically, this structure yields super-exponential suppression of the LB-weighted stationary density and a vanishing probability current at the origin. The behaviour at large scale factor is sector dependent: in de Sitter regimes, $a=\infty$ is a regular boundary with $P_*(a)\propto a^{-2}$ and $\mathcal P_*(a)\propto a^{-3/2}$, while sufficiently confining dynamics or $\Lambda<0$ lead to $a=\infty$ becoming an entrance boundary and a globally normalizable stationary state. Therefore, while the avoidance of the classical singularity at $a=0$ emerges from the stochastic dynamics, the infrared completion of the probability measure is controlled by the underlying cosmological sector.

This latter mechanism stands in contrast with structural stochastic proposals such as composite stochastic gravity, where diffusion is postulated externally~\cite{erlich2018stochastic, erlich2022first}. In the present framework, the multiplicative structure $\sigma(a)\propto a^2$ emerges directly from FLRW geometry and LB quantisation, making the boundary classification a direct consequence of minisuperspace backreaction rather than an external assumption. Accordingly, the stochastic trajectories obtained in the semiclassical limit are not additional ad hoc elements but arise as the effective dynamical description of the reduced, open quantum system \cite{paz1993environment}, with the WD wavefunction playing the role of a generator of statistical structure and correlations rather than representing a single ontic cosmological history. That is, $\Psi$ should be interpreted as encoding constraints, decoherence structure, and ensemble weights for possible minisuperspace realisations, while the physically realised universe corresponds to a single stochastic trajectory sampled from this reduced dynamics.

We emphasize that the present framework applies to homogeneous minisuperspace and employs the Markov approximation to the influence kernel. Future developments include: (i) extending the construction to anisotropic Bianchi models to test the robustness of the entrance boundary, (ii) incorporating non-Markovian influence kernels with explicit memory, and (iii) embedding the formalism in stochastic-inflation frameworks to investigate potential observational signatures~\cite{vennin2015correlation}. Together, these would determine whether the stochastic backreaction mechanism identified here remains consistent beyond the minisuperspace limits and, importantly, whether it provides a controlled route to singularity resolution in quantum cosmology.

\appendix
\section{Stochastic Wheeler–DeWitt Dynamics: Derivation and Noise Scaling}
\label{app:appA}
\renewcommand{\theequation}{A\arabic{equation}}
\setcounter{equation}{0}

In this appendix, we summarize the canonical minisuperspace reduction and relational quantization, derive the stochastic backreaction induced by coarse-grained quantum fluctuations using an open-system approach, and analyze the resulting semiclassical stochastic dynamics. We also examine the robustness of the diffusion structure and its implications for the behavior of the minisuperspace dynamics near the classical singularity.
\subsection{Classical Minisuperspace Reduction}
\label{app:a1}
We begin from the ADM action for gravity minimally coupled to a scalar field,
\begin{equation}
S[g,\phi] = \frac{1}{16\pi G}\int d^4x\,\sqrt{-g}\,(R-2\Lambda) 
- \int d^4x\,\sqrt{-g}\,\Big[\tfrac12 g^{\mu\nu}\partial_\mu\phi\,\partial_\nu\phi + V(\phi)\Big].
\label{eq:fullaction}
\end{equation}
Imposing spatial homogeneity and isotropy, the spacetime metric takes the FLRW form
\begin{equation}
ds^2 = -N^2(t)\,dt^2 + a^2(t)\,\gamma_{ij}dx^i dx^j,
\end{equation}
where $N(t)$ is the lapse function, $a(t)$ the scale factor, and $\gamma_{ij}$ the metric of a maximally symmetric spatial slice with curvature $k\in\{-1,0,+1\}$. Performing the minisuperspace reduction yields
\begin{equation}
S_{\mathrm{ms}}[a,\varphi] = 2\pi^2\int dt\,N \left[-\frac{3a\dot{a}^2}{N^2} + 3ka - \Lambda a^3 
+ \frac{a^3\dot{\varphi}^2}{2N^2} - a^3 V(\varphi)\right],
\label{eq:msaction}
\end{equation}
where $\varphi(t)$ denotes the homogeneous scalar mode, commonly used in quantum cosmology \cite{david2010consistent, paliathanasis2024minisuperspace, halliwell1989decoherence}. The associated canonical momenta are \(p_a = -12\pi^2\frac{a\dot{a}}{N}\) and \(p_\varphi = 2\pi^2\frac{a^3\dot{\varphi}}{N},\) and the Hamiltonian constraint takes the form
\begin{equation}
H_0(a,p_a,\varphi,p_\varphi) 
= -\,\frac{p_a^2}{24\pi^2 a} - 2\pi^2(3ka-\Lambda a^3)
+ \frac{p_\varphi^2}{4\pi^2 a^3} + 2\pi^2 a^3 V(\varphi)=0.
\label{eq:H0}
\end{equation}

Fixing the homogeneous scalar field as a relational clock by identifying its value with a relational time parameter, $\varphi(t)\equiv \phi$, and solving \eqref{eq:H0} for $p_\varphi$ yields the effective Hamiltonian generating evolution in the clock variable $\phi$, which is Eq. \eqref{eq:Heff} in the main text.

\subsection{System--Environment Coupling}
\label{app:a2}
We now reintroduce perturbative degrees of freedom and treat the homogeneous $(a,\varphi)$ as a system coupled to an environment. Decomposing the metric and scalar field as
\begin{equation}
g_{\mu\nu} = g^{(0)}_{\mu\nu}[a] + h_{\mu\nu}, \qquad 
\phi = \varphi + \delta\phi,
\end{equation}
and expanding the action (\ref{eq:fullaction}) to quadratic order in the perturbations $(h_{\mu\nu},\delta\phi)$ yields
\begin{equation}
S[g,\phi] = S_{\mathrm{ms}}[a,\varphi] 
+ \frac{1}{2}\int d^4x\,\chi^T \mathcal{D}[a,\varphi]\chi 
+ \int d^4x\,\mathcal{I}[a,\varphi;\chi],
\label{eq:splitS}
\end{equation}
where $\chi$ collectively denotes the gauge-fixed perturbation fields (TT tensor modes and gauge-invariant scalar modes), $\mathcal{D}$ is the quadratic fluctuation operator, and $\mathcal{I}$ contains the terms linear in $\chi$ which encode the system--environment coupling. For minimally coupled matter, the only explicit dependence on the background scale factor in the potential sector comes from the volume element $\sqrt{-g}\sim a^3$, so that the scalar potential contribution to the interaction term is of the form
\begin{equation}
S_{\mathrm{int}} = -\int dt\, a^3(t)V(\varphi(t)) \int d^3x\,\mathcal{J}(t,\mathbf{x}),
\end{equation}
where $\mathcal{J}$ is a linear functional of the perturbations. After integrating over the comoving spatial volume, the system--bath coupling reduces to
\begin{equation}
\int dt\, O_{\mathrm{raw}}(a,\phi)\,\mathcal{B}(t), \qquad 
O_{\mathrm{raw}}(a,\phi) = \alpha\,a^3 V(\phi),
\label{eq:Oral}
\end{equation}
where $\mathcal{B}(t)$ is a suitably normalized collective bath operator and $\alpha$ is a constant absorbing geometrical factors. We emphasize that $O_{\mathrm{raw}}(a,\phi)\propto a^3$ follows purely from geometry and minimal coupling and is independent of the details of the perturbation spectrum.

Treating $\chi$ as an environment and tracing it out in the Schwinger--Keldysh formalism yields the Feynman--Vernon influence functional \cite{feynman2000theory},
\begin{equation}
\mathcal{F}[a_+,a_-] = \exp\Big( i S_{\mathrm{IF}}[a_+,a_-] \Big),
\end{equation}
with the influence phase
\begin{equation}
S_{\mathrm{IF}}[a_+,a_-] 
= \frac{1}{2}\int dt\,dt'\,\Big[O_{\mathrm{raw}}(t)\,G_R(t,t')\,O_{\mathrm{raw}}(t') 
+ i\, O_{\mathrm{raw}}(t)\,N(t,t')\,O_{\mathrm{raw}}(t') \Big].
\label{eq:SIF}
\end{equation}
Here $G_R$ is a retarded kernel encoding dissipation, and $N$ is the Hadamard (symmetric) noise kernel. Their explicit expressions can be written in terms of the environment correlation functions as
\begin{equation}
G_R(t,t') = i\,\theta(t-t')\langle [\mathcal{B}(t),\mathcal{B}(t')]\rangle, \qquad
N(t,t') = \tfrac12\langle \{\mathcal{B}(t),\mathcal{B}(t')\}\rangle,
\end{equation}
with expectation values taken in a chosen perturbative quantum state. The imaginary part of $S_{\mathrm{IF}}$ can be represented by a classical Gaussian stochastic field $\xi(t)$ via a Hubbard--Stratonovich transformation,
\begin{equation}
\exp\Big(-\frac{1}{2}\int dt\,dt'\,O_{\mathrm{raw}}(t)N(t,t')O_{\mathrm{raw}}(t')\Big)
= \int \mathcal{D}\xi\,\mathcal{P}[\xi]\,
\exp\Big(i\int dt\,O_{\mathrm{raw}}(t)\,\xi(t)\Big),
\end{equation}
with
\begin{equation}
\langle\xi(t)\rangle = 0, \qquad \langle\xi(t)\xi(t')\rangle = N(t,t').
\label{eq:xicorr}
\end{equation}

\subsection{Semiclassical Limit and Stochastic Hamilton--Jacobi Dynamics}
\label{app:a3}
It follows that the reduced minisuperspace wave functional obeys a stochastic Schr\"odinger equation \cite{calzetta1994noise, calzetta2009nonequilibrium},
\begin{equation}
i\hbar\,\partial_\phi \Psi(a,\phi;\xi) 
= \big(\hat{H}_{\mathrm{eff}}(a,\phi) + \xi(\phi)\,\hat{O}_{\mathrm{raw}}(a,\phi)\big)\Psi(a,\phi;\xi),
\label{eq:sSE}
\end{equation}
where $\hat{H}_{\mathrm{eff}}$ is the LB-ordered Hamiltonian derived from (\ref{eq:Heff}) and $\hat{O}_{\mathrm{raw}}(a,\phi)$ is the operator corresponding to $O_{\mathrm{raw}}(a,\phi)=\alpha a^3 V(\phi)$. To obtain a semiclassical stochastic dynamics for the expectation trajectory of $a(\phi)$, we adopt the usual WKB ansatz $\Psi(a,\phi;\xi)=A(a,\phi;\xi)\exp[iS(a,\phi;\xi)/\hbar]$ and retain only terms of order $\hbar^0$ in (\ref{eq:sSE}). This results in a stochastic Hamilton--Jacobi equation
\begin{equation}
\partial_\phi S(a,\phi;\xi) + H_{\mathrm{eff}}\big(a,\partial_a S;\phi\big) + \xi(\phi)\,O_{\mathrm{raw}}(a,\phi) = 0.
\label{eq:stochHJ}
\end{equation} Defining $p_a(a,\phi;\xi) := \partial_a S(a,\phi;\xi)$, Eq.~(\ref{eq:stochHJ}) yields stochastic canonical equations
\begin{equation}
\frac{da}{d\phi} = \frac{\partial H_{\mathrm{eff}}}{\partial p_a}(a,p_a;\phi), \qquad
\frac{dp_a}{d\phi} = -\frac{\partial H_{\mathrm{eff}}}{\partial a}(a,p_a;\phi) - \xi(\phi)\,\partial_a O_{\mathrm{raw}}(a,\phi),
\label{eq:canonSDE}
\end{equation}
with $\xi(\phi)$ a stationary Gaussian process with covariance \eqref{eq:xicorr}. The explicit form of $\partial H_{\mathrm{eff}}/\partial p_a$ depends on the detailed balance among the kinetic, curvature, cosmological constant and potential terms in (\ref{eq:Heff}), but near $a\to0$ and for regular $V(\phi)$ and $\Lambda$, the classical drift term can be shown to behave as $v(a) = c a + O(a^3)$ with $c>0$, as discussed in the main text. The key point is that the stochastic forcing in $p_a$ is proportional to $\partial_a O_{\mathrm{raw}}$,
\begin{equation}
\Xi(a,\phi;\xi) := -\xi(\phi)\,\partial_a O_{\mathrm{raw}}(a,\phi)
= -\alpha\,\xi(\phi)\,\partial_a\big[a^3 V(\phi)\big] 
= -3\alpha\,V(\phi)\,a^2\,\xi(\phi),
\label{eq:Xi}
\end{equation}
so that the noise amplitude in the momentum equation scales as $a^2$ for small $a$.

Equation \eqref{eq:canonSDE} must be translated into an effective SDE for $a(\phi)$ alone. In general, for a two-dimensional Hamiltonian system, it is not possible to eliminate $p_a$ exactly. However, in the semiclassical regime relevant here, one can adopt a narrow-tube approximation in which the stochastic process remains close to a classical trajectory in phase space~\cite{calzetta2009nonequilibrium}, so that $p_a$ can be expressed as a regular function of $a$ and $\phi$ up to small stochastic deviations. The stochastic correction to $\dot a := da/d\phi$ inherits the same multiplicative scaling as $\Xi(a,\phi;\xi)$ up to factors that are regular and non-vanishing at $a=0$. Thus, we may write the effective one-dimensional Stratonovich SDE
\begin{equation}
\frac{da}{d\phi} = v(a) + \sigma(a)\circ \eta(\phi),
\label{eq:effSDE}
\end{equation}
where $\eta(\phi)$ is normalized white noise with $\langle \eta(\phi)\eta(\phi')\rangle = \delta(\phi-\phi')$, and the prefactor is of the form
\begin{equation}
\sigma(a) = \sqrt{\zeta}\,a^2 + O(a^2),
\label{eq:sigma_a2}
\end{equation}
for $\zeta>0$ determined by the noise kernel $N$ and the coupling strength $\alpha$. The $O(a^2)$ term denotes corrections that are subleading in the limit $a\to0$ and do not modify the leading exponent. This leads to the FP equation for the flat-measure density $P(a,\phi)$
\begin{equation}
\partial_\phi P(a,\phi) = -\partial_a\big[f(a)P(a,\phi)\big] + \frac{1}{2}\partial_a^2\big[\sigma^2(a)P(a,\phi)\big],
\end{equation}
with drift $f(a)=v(a)+\frac{1}{2}\sigma(a)\sigma'(a)$ and diffusion coefficient $D(a)=\frac{1}{2}\sigma^2(a)\propto a^4$ at leading order, as used in the main text. The LB measure $\mu(a)\propto a^{1/2}$ then defines the physical probability density $\mathcal{P}(a,\phi)=\mu(a)P(a,\phi)$ without affecting the leading small-$a$ exponent.
\subsection{Boundary Behavior and Power-Law Classification of Diffusion}
\label{app:a4}
To assess whether other small-$a$ behaviours of $\sigma(a)$ are compatible with the above microscopic picture, it is convenient to generalize the analysis by considering the power-law ansatz
\begin{equation}
\sigma(a) \sim a^p, \qquad a\to0,
\label{eq:sigmap}
\end{equation}
with $p\in\mathbb{R}$ a priori free. This encompasses both the derived case $p=2$ and potential alternatives that might arise from non-minimal couplings or non-Markovian noise kernels. The Feller classification of the boundary at $a=0$ is encoded in the scale function $s(a)$ and the speed density $m(a)$,
\begin{equation}
s(a) = \exp\left\{-\int^a \frac{2f(x)}{\sigma^2(x)}\,dx\right\}, \qquad
m(a) = \frac{1}{\sigma^2(a)s(a)},
\label{eq:scale_speed}
\end{equation}
and the integrability properties of $s$ and $m$ near the boundary. Assuming the drift retains the linear behaviour $f(a) = c a + O(a^3)$ with $c>0$ and inserting (\ref{eq:sigmap}) yields
\begin{equation}
\frac{2f(a)}{\sigma^2(a)} \sim \frac{2c a}{\zeta a^{2p}} = \frac{2c}{\zeta} a^{1-2p}.
\end{equation}
For $p\neq 1$, integration gives
\begin{equation}
\int^a \frac{2f(x)}{\sigma^2(x)}\,dx \sim \frac{2c}{\zeta(2-2p)}\,a^{2-2p} + C,
\end{equation}
so that
\begin{equation}
s(a) \sim C_1 \exp\left[-\frac{2c}{\zeta(2-2p)}\,a^{2-2p}\right], \qquad 
m(a)\sim \frac{1}{\zeta C_1} a^{-2p}\exp\left[\frac{2c}{\zeta(2-2p)}\,a^{2-2p}\right],
\end{equation}
with constants $C,C_1>0$. For $p>1$ we have $2-2p<0$ and $a^{2-2p}\to\infty$ as $a\to0$, so the exponent in $s(a)$ is positive and diverging, implying $s(a)\to\infty$ and
\begin{equation}
\int_0^\varepsilon s(a)\,da = +\infty.
\end{equation}
At the same time, the exponential in $m(a)$ decays faster than any power as $a\to0$, making $m(a)$ integrable in a neighbourhood of zero. Thus, $a=0$ is an entrance boundary for all $p>1$, and trajectories started at $a>0$ almost surely never reach the singularity in finite relational time. For $p<1$, the exponent in $s(a)$ tends to a finite constant as $a\to0$, so $s(a)$ remains finite and $\int_0^\varepsilon s(a)\,da<\infty$. In that case, the boundary is accessible and, generically, attainable. The marginal case $p=1$ requires a separate analysis: one finds $s(a)\sim a^{-2c/\zeta}$ and $m(a)\sim a^{-2+2c/\zeta}$, implying that $a=0$ is entrance only if $c/\zeta>1/2$; otherwise it is not. 
\subsection{Comments on Microscopic Constraints}

We now ask which values of $p$ can be generated by the microscopic structure of the influence functional. Suppose more generally that the system--environment coupling takes the form
\begin{equation}
O_{\mathrm{raw}}(a,\phi) = \alpha\,a^q\,\mathcal{U}(\phi),
\label{eq:Oq}
\end{equation}
with $q\in\mathbb{R}$ and a bounded function $\mathcal{U}(\phi)$, thereby allowing for non-minimal or curvature-dependent couplings. The stochastic force in the momentum equation (\ref{eq:canonSDE}) then scales as
\begin{equation}
\Xi(a,\phi;\xi) = -\xi(\phi)\,\partial_a O_{\mathrm{raw}}(a,\phi)
\sim -\alpha q\,\mathcal{U}(\phi)\,a^{q-1}\,\xi(\phi),
\end{equation}
so that, in the narrow-tube approximation, the multiplicative amplitude entering the effective SDE for $a$ scales as
\begin{equation}
\sigma(a) \sim a^{q-1},
\end{equation}
up to factors that are regular and non-vanishing at $a=0$. This identifies the exponent $p$ in \eqref{eq:sigmap} as $p=q-1$. For minimally coupled gravity and scalar field, we have argued above that $O_{\mathrm{raw}}\propto a^3$, i.e. $q=3$, implying $p=2$. By contrast, $q<3$ would require the removal of at least one power of $a$ from the volume factor, which is not compatible with the local form of $\sqrt{-g}$, while $q>3$ would require additional powers of $a$ from curvature invariants or higher-derivative couplings, which can only arise from higher-order counterterms or non-minimal couplings at Planckian energies. Consequently, within the semiclassical regime and for minimally coupled matter, the choice $q=3$ and hence $p=2$ is uniquely selected.

Finally, we comment on possible non-power-law corrections. One-loop nonlocal effective actions in curved spacetime \cite{barvinsky1990covariant, barvinsky1995one} can generate logarithmic modifications to the effective couplings, leading to
\begin{equation}
\sigma(a) \sim a^2\big(1 + \beta\log(a/a_0)\big),
\end{equation}
while asymptotic silence or polymer quantization scenarios may produce exponentially suppressed forms $\sigma(a)\sim a^2\exp(-\alpha/a^r)$ at extremely small $a$. In both cases, the leading small-$a$ behaviour remains $a^2$ for all $a$ in the semiclassical domain, with the logarithmic factors affecting only the prefactor $\zeta$ or becoming relevant only in a regime where the minisuperspace truncation is no longer reliable. As such, these corrections do not represent genuinely distinct diffusion structures in the context considered here.

Collecting these results, we emphasize that the multiplicative diffusion coefficient in the sWD minisuperspace dynamics has the asymptotic form
\begin{equation}
\sigma(a) = \sqrt{\zeta}\,a^2 + O\big(a^2\log a\big), \qquad a\to0,
\end{equation}
with the leading exponent $p=2$ fixed by the combination of (i) geometric minimal coupling, (ii) the canonical structure of the FLRW minisuperspace, (iii) Laplace--Beltrami quantization, and (iv) the requirement of a dynamically inaccessible boundary at $a=0$ without fine-tuning. As a result, the specific form presented in the main text represents the most natural (choice of) scaling consistent with the underlying semiclassical framework.

\section{Noise kernel and Markov limit from coarse-grained gravitons}
\label{app:B_markov}
\renewcommand{\theequation}{B\arabic{equation}}

Here, we illustrate in a simple setting how a local (white) noise kernel arises when the minisuperspace degrees of freedom couple to coarse-grained TT graviton modes, and under which conditions dissipative effects are subleading in the minisuperspace dynamics. This appendix is not meant to provide a fully realistic model of cosmological perturbations, but rather to substantiate the Markov approximation used in the main text and in Appendix~\ref{app:a3} and~\ref{app:a4}.

We consider small TT perturbations $h_{ij}^{\rm TT}$ around the FLRW background,
\begin{equation}
g_{ij}(t,\mathbf{x}) = a^2(t)\Big[\gamma_{ij} + h_{ij}^{\rm TT}(t,\mathbf{x})\Big],
\end{equation}
with $h_{ij}^{\rm TT}$ transverse and traceless with respect to the spatial metric $\gamma_{ij}$. At quadratic order, the graviton sector can be written as a set of decoupled harmonic oscillators with time-dependent frequencies \cite{hu1990quantum}. For our purposes, it suffices to use the mode expansion
\begin{equation}
h_{ij}^{\rm TT}(t,\mathbf{x})
= \sum_{\lambda=+,\times}\int\!\frac{d^3k}{(2\pi)^3}\,
\epsilon^{(\lambda)}_{ij}(\mathbf{k})\,h_{\mathbf{k}}^{(\lambda)}(t)\,
e^{i\mathbf{k}\cdot\mathbf{x}},
\end{equation}
where $\epsilon^{(\lambda)}_{ij}$ are the usual TT polarisation tensors and $h_{\mathbf{k}}^{(\lambda)}(t)$ are the mode amplitudes. Introducing rescaled variables $\chi_{\mathbf{k}}^{(\lambda)}$ that obey approximately free oscillator equations, the graviton Hamiltonian reads schematically
\begin{equation}
H_{\rm grav} \simeq \frac{1}{2}\sum_{\lambda}\int d^3k\,
\big[\Pi_{\mathbf{k}}^{(\lambda)}\Pi_{-\mathbf{k}}^{(\lambda)}
+\omega_k^2(t)\,\chi_{\mathbf{k}}^{(\lambda)}\chi_{-\mathbf{k}}^{(\lambda)}\big],
\end{equation}
with $\Pi_{\mathbf{k}}^{(\lambda)}$ the canonical momenta and $\omega_k(t)\simeq k/a(t)$ the instantaneous frequencies for subhorizon modes. In this approximation, the TT graviton sector represents a Gaussian environment for the minisuperspace variable $a$, with a large set of independent oscillators labelled by $(\mathbf{k},\lambda)$.

As in Appendix~\ref{app:appA}, we parametrise the system–bath coupling in the influence functional by
\begin{equation}
S_{\rm int} = \int d\phi\,O_{\rm raw}(a,\phi)\,\mathcal{B}(\phi),
\end{equation}
where $O_{\rm raw}(a,\phi)=\alpha\,a^3 V(\phi)$ carries all the $a$-dependence and $\mathcal{B}(\phi)$ is a linear combination of TT modes evaluated along the minisuperspace trajectory. For definiteness, we may take
\begin{equation}
\mathcal{B}(\phi) = \sum_{\lambda}\int_{|\mathbf{k}|<k_c}\!\frac{d^3k}{(2\pi)^3}\,
c_k^{(\lambda)}\,\chi_{\mathbf{k}}^{(\lambda)}(\phi),
\label{eq:Boperator}
\end{equation}
where the coefficients $c_k^{(\lambda)}$ encode the projection onto the minisuperspace variable and the integral is restricted to comoving momenta below a coarse-graining cutoff $k_c$. The restriction $|\mathbf{k}|<k_c$ is the origin of the finite correlation time of the effective noise kernel.

Assuming the environment is in a Gaussian adiabatic state, such as the Bunch–Davies vacuum in a slowly varying background \cite{bunch1978quantum}, the TT mode functions can be written as
\begin{equation}
\chi_{\mathbf{k}}^{(\lambda)}(\phi)
= u_k(\phi)\,a_{\mathbf{k}}^{(\lambda)} + u_k^\ast(\phi)\,a_{-\mathbf{k}}^{(\lambda)\dagger},
\end{equation}
where $a_{\mathbf{k}}^{(\lambda)}$ and $a_{\mathbf{k}}^{(\lambda)\dagger}$ are annihilation and creation operators, and the mode functions are well approximated by
\begin{equation}
u_k(\phi)\simeq \frac{1}{\sqrt{2\omega_k}}\,e^{-i\omega_k(\phi-\phi_0)}, \qquad \omega_k\simeq \frac{k}{a(\phi)},
\label{eq:modefunc}
\end{equation}
whenever the minisuperspace variables vary slowly on the timescale set by $1/\omega_k$. Inserting \eqref{eq:Boperator} and \eqref{eq:modefunc} into the definition of the Hadamard (noise) kernel,
\begin{equation}
N(\phi,\phi') = \frac{1}{2}\,\big\langle\{\mathcal{B}(\phi),\mathcal{B}(\phi')\}\big\rangle,
\end{equation}
and using $\langle a_{\mathbf{k}}^{(\lambda)} a_{\mathbf{k}'}^{(\lambda')\dagger}\rangle=(2\pi)^3\delta^{(3)}(\mathbf{k}-\mathbf{k}')\delta_{\lambda\lambda'}$, we obtain
\begin{equation}
N(\phi,\phi')
\simeq \int_{|\mathbf{k}|<k_c}\!\frac{d^3k}{(2\pi)^3}\,
|c_k|^2\,\frac{\cos\big[\omega_k(\phi-\phi')\big]}{2\omega_k},
\label{eq:Nphi}
\end{equation}
where we have suppressed the polarisation label and absorbed it into $|c_k|^2$. For massless modes and slowly varying $a(\phi)$, $\omega_k\simeq k/a(\phi)$ varies adiabatically, and for time separations $\Delta\phi:=\phi-\phi'$ much smaller than the clock timescale $\tau_{\rm clock}$ it is a good approximation to treat $\omega_k$ as constant. Equation \eqref{eq:Nphi} then takes the form of a standard ultraviolet-regularised cosine transform,
\begin{equation}
N(\Delta\phi)
\simeq \int_0^{k_c}\!\frac{4\pi k^2 dk}{(2\pi)^3}\,
\frac{|c_k|^2}{2\omega_k}\,\cos(\omega_k\Delta\phi).
\end{equation}
If the projection coefficients $|c_k|^2/(2\omega_k)$ vary mildly with $k$ for $k<k_c$, the dominant dependence on $\Delta\phi$ arises from the oscillatory factor. In the limit of a large cutoff $k_c$ and for test functions $F(\phi)$ that vary slowly on timescales $\Delta\phi\gtrsim \tau_{\rm clock}$, the integral over $k$ can be approximated via standard arguments of the Riemann–Lebesgue lemma and stationary phase as a representation of the Dirac delta,
\begin{equation}
\int d\phi'\,N(\phi-\phi')\,F(\phi')
\simeq \zeta\,F(\phi),\qquad
N(\phi,\phi')\simeq 2\zeta\,\delta(\phi-\phi'),
\label{eq:Markov}
\end{equation}
where the effective diffusion strength is given by the spectral integral
\begin{equation}
\zeta \sim \int_0^{k_c}\!\frac{4\pi k^2 dk}{(2\pi)^3}\,
\frac{|c_k|^2}{4\omega_k}.
\end{equation}
The Markov limit \eqref{eq:Markov} is therefore valid provided that the graviton correlation time $\tau_{\rm grav}\sim 1/k_c$ is much shorter than the minisuperspace evolution time $\tau_{\rm clock}$, and that the spectral weight $|c_k|^2/(2\omega_k)$ does not introduce sharp additional structure on scales comparable to $\tau_{\rm clock}$. 

The corresponding retarded kernel is
\begin{equation}
G_R(\phi,\phi') = i\,\theta(\phi-\phi')\,\big\langle[\mathcal{B}(\phi),\mathcal{B}(\phi')]\big\rangle
\simeq \int_0^{k_c}\!\frac{4\pi k^2 dk}{(2\pi)^3}\,
|c_k|^2\,\frac{\sin[\omega_k(\phi-\phi')]}{2\omega_k},
\end{equation}
and is related to the noise kernel through the fluctuation–dissipation relation in frequency space \cite{calzetta1994noise}. In the high-frequency regime, the spectral weight of $\mathrm{Im}\,G_R(\omega)$ is concentrated at $\omega\sim\omega_k\gg 1/\tau_{\rm clock}$, whereas the minisuperspace dynamics is confined to much lower effective frequencies. Consequently, the dissipative response is parametrically suppressed at the level of the relational evolution equation and contributes only a small renormalization of the deterministic drift. To leading order in the separation of scales, integrating out coarse-grained gravitons therefore produces a local stochastic force of strength $\zeta$, while dissipation and weakly coloured corrections remain subdominant and do not affect the entrance character of the boundary at $a\to0$. A fully non-Markovian treatment, in which long-range memory may influence the infrared behaviour, is beyond the scope of this Letter.

\printbibliography

\end{document}